\documentclass[aps,showpacs,preprint]{revtex4}
\usepackage{graphicx}
\begin{document}
\title{Effect of system level structure and spectral distribution of the environment on the decoherence rate
\footnote{Corresponding authors:
Jingfu Zhang, zhangjfu2000@yahoo.com, Jingfu@e3.physik.uni-dortmund.de;\\
Dieter Suter, Dieter.Suter@uni-dortmund.de }}
\author{Jingfu Zhang, Xinhua Peng, Nageswaran Rajendran, and Dieter Suter}
\address{Fachbereich Physik, Universit$\ddot{a}$t Dortmund, 44221 Dortmund, Germany\\
}
\date{\today}

\begin{abstract}

Minimizing the effect of decoherence on a quantum register must be a
central part of any strategy to realize scalable quantum information
processing. Apart from the strength of the coupling to the
environment, the decoherence rate is determined by the the system
level structure and by the spectral composition of the noise trace
that the environment generates. Here, we discuss a relatively simple
model that allows us to study these different effects quantitatively
in detail. We evaluate the effect that the perturbation has on a
nuclear magnetic resonance (NMR) system while it performs a Grover
search algorithm.
\end{abstract}

\maketitle
\section{Introduction}
The advantage that quantum computers have over classical computers
hinges on the creation and preservation of quantum
coherence \cite{Unruh}. Any real quantum computer, however,
interacts with its environment and such interactions result in
decoherence which increases the probability that the quantum computation
 may fail \cite{Chuang}.
 Decoherence is thus one of the main obstacles for building
practical quantum computers. 

 A number of strategies have developed for suppressing decoherence,
including quantum error corrections \cite{shor95}, dynamical
decoupling (quantum control) \cite{Viola}, decoherence-free
subspaces \cite{df}, holonomic quantum computation \cite{holo}, the
quantum Zeno effect \cite{zeno}, and spectral degeneracy systems
\cite{Grigorenko,Grigorenko2}. It has been proved that the first
four strategies can be unified under a general algebraic framework
\cite{Knill}, and the quantum Zeno effect can be unified with
dynamical decoupling \cite{Facchi}. Up to the present quantum error
corrections \cite{qec}, decoherence-free subspaces \cite{free},
holonomic quantum computation \cite{gc}, and the suppression of
artificial decoherence by dynamical decoupling (bang-bang control)
\cite{Kondo} have been experimentally tested using nuclear magnetic
resonance (NMR).

In this paper, we concentrate on a different aspect:
The decoherence rate is not only determined by the strength of the
coupling to the environment, but also by its operational form.
As an example, the decoherence differs qualitatively if the coupling
operator commutes with the system operator.
Another aspect is the spectral composition of the noise:
If the environment is (almost) static, the interaction is adiabatic.
If it has components that fluctuate at transition frequencies of the
system Hamiltonian, its effect can be particularly strong.

A number of model systems have been discussed to study the
interaction of a quantum register with a noisy environment. One
model is known as the spin bath where the environment consists of a
set of two- level systems or spin-1/2 systems \cite{spinbath}. In
another model, the so-called spin-boson model, the environment
consists of a set of harmonic oscillators
\cite{Wilhelm,Srorcz,Srorcz2,Romero}. NMR can simulate the
decoherence effect (or called artificial decoherence) through the
interactions generated by the spins viewed as spin bath
\cite{Kondo,Teklemariam,errmodel} or implement error models by
radio- frequency and gradient pulses for demonstrating quantum error
corrections \cite{qec}.

Here, we use a semiclassical model, where the environment acts on
the system through classical stochastic fields. The effect on the
system is the same as that of other environments, provided a
suitable ensemble average is taken. The coupling to the system
occurs through spin operators. We distinguish two systems, in one of
which the coupling operator commutes with the system Hamiltonian, in
the other it does not. For both systems, we implement a Grover
search algorithm\cite{Grover,realize} and demonstrate the effect of
different environments.

\section{System and Environment}

\subsection{System Hamiltonian}

We use a system of two qubits to compare the effect of different
colored noise for different energy level structures.
The Hamiltonian of the system is
\begin{equation}\label{Hrs} 
   H_{s}=\frac{1}{2}\hbar \left[ \omega^{1}_{z}\sigma_{z}^{1}
   + \omega^{2}_{z}\sigma_{z}^{2}
   - \omega^{2}_{x}\sigma_{x}^{2}+
  \pi J\sigma_{z}^{1}\sigma_{z}^{2} \right] .
\end{equation}
Here $\sigma^{i}_{x/z}$ denotes the $x$ or $z$ component of the
Pauli matrix for spin $i$, $\omega_{x/z}$ describes the strength
of the magnetic field along the $x$ or $z$ axis, respectively, and $J$
denotes the coupling constant.

For suitable parameter sets, this Hamiltonian can execute a
controlled NOT (CNOT) operation in a single step, without external
control operations \cite{Grigorenko2}; for a wider set of
parameters, two-qubit gate operations can be executed that fall into
the CNOT equivalence class, i.e. they are equivalent to the CNOT
operation up to single qubit operations. Table \ref{t:pars}
summarizes two parameter sets: System {\rm I} ( left hand column )
gives the parameters for the CNOT operation, system {\rm II} ( right
hand column) the parameters for a CNOT-equivalent operation. We
write the Hamiltonians of the 2 systems $H_{s}^{\rm I}$ and
$H_{s}^{\rm II}$. Their eigenvalues are also shown in Table
\ref{t:pars}. The transition angular frequencies are
$\omega_{nm}=(E_{n}-E_{m})/\hbar$ for each system, where $n$, $m=1,
2, 3, 4$.
The different energy
level structures of the two systems provides the possibility to suppress
decoherence induced by the coupling to the environment.

\begin{table}[htdp]
\caption{Parameters for the two systems in units of $\pi J$, where $J$ is the coupling
constant.}
\begin{center}
\begin{tabular}{|c|c|c|}
\hline
 & System {\rm I} & System  {\rm II} \\
\hline $\omega_{z}^{1}$ & $0.378$  & $0.378$  \\
\hline $\omega^{2}_{z}$  & 1 $$      & 1  $$    \\
\hline $\omega^{2}_{x}$  & $2.272$ & $1.136$ \\
\hline $E_1$  & $-1.32\hbar$ & $-0.961\hbar$\\
\hline $E_2$  & $-1.32\hbar$ & $-0.758\hbar$\\
\hline $E_3$  & $0.948\hbar$ & $0.379\hbar$\\
\hline $E_4$  & $1.70\hbar$ & $1.34\hbar$\\
\hline
\end{tabular}
\end{center}
\label{t:pars}
\end{table}

\subsection{Coupling to the Bath} \label{sect2}

We now consider decoherence processes that are induced by a coupling
to the environment that takes the form
\begin{equation}\label{Hen}
    H(t)=H_{s}+\hbar\pi s(t)A .
\end{equation}
For our purpose, the system operator $A$ may be either
$\sigma_{z}^{1}$, $\sigma_{z}^{2}$, or
$\sigma_{z}^{1}+\sigma_{z}^{2}$. The bath term $s(t)$ of the
coupling operator may be either a classical random field or a
quantum mechanical operator; for our purposes, it will be
sufficient to consider it a time-dependent magnetic field with
zero mean.

For the random perturbation $s(t)$, we consider stochastic functions
with a Lorentzian spectral distribution of the power spectrum
\begin{equation}\label{Lor}
S(\omega)=\frac{\kappa\Gamma}{\Gamma^{2}+(\omega-\omega_{0})^2}
\end{equation}
and check the effect of the center frequency $\omega_0$ on the
decoherence rate in the two systems, where $\kappa$ describes the
strength of $S(\omega)$.

\begin{figure}
\includegraphics[width=5in]{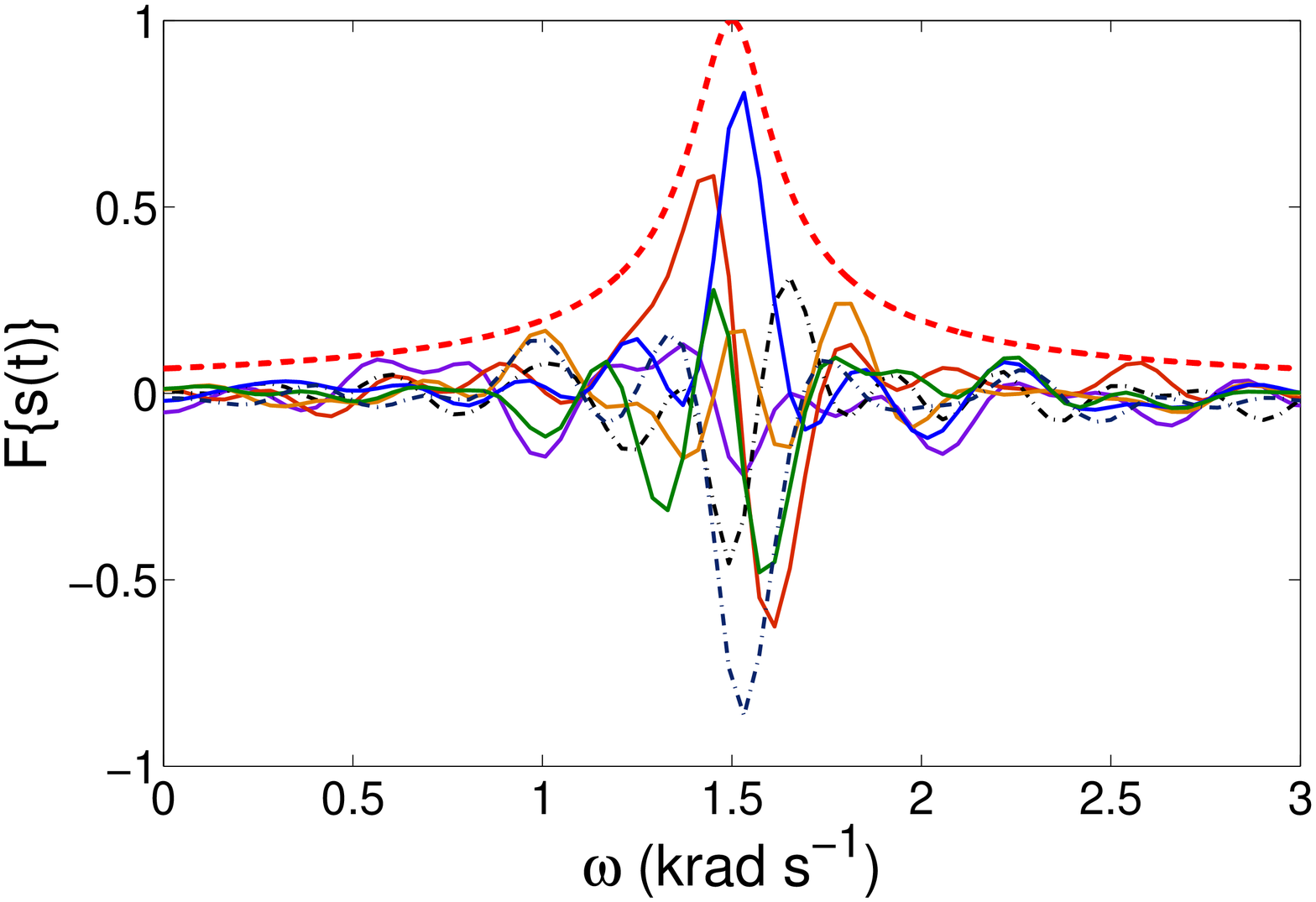}    
\caption{(Color online) Spectra (solid and dot dashed curves) of
some random functions $s(t)$. The dashed curve represents the
envelope function $\sqrt{S(\omega)}$.} \label{figs}
\end{figure}

In the experiment, the coupling constant $J$ had the value $J=215$
s$^{-1}$. To simulate the effect of the environment, we generated
the stochastic functions by digital filtering of a random time
series. Each time series had a duration of $24.35$ ms consisted of
$80$ segments. For each time series we performed an experiment and
summed over the individual experimental data. Figure \ref{figs}
shows the power spectra of some time series $s(t)$ and the RMS
spectral density $\sqrt{S(\omega)}$. Figure 2 shows the spectral
density functions \cite{Abragam} for the four reservoirs that we
compare, in relation to the transition frequencies of the system.

\section{Grover search in noisy systems }

   The elementary gates in the Grover search are the Walsh- Hadamard
 transform  and the controlled phase
reversal $I_{|x\rangle}$ where $|x\rangle$ denotes a computational
basis state, e.g. $|00\rangle$ or $|11 \rangle$. $I_{|x\rangle}$ can
be implemented by CNOT gates and one- qubit operations
\cite{Srorcz2}. Hence we first implement the CNOT gate using the
evolution under $H_s$ (\ref{Hrs}).

\subsection{Implementation of CNOT gates}

As stated in the introduction, the system Hamiltonian generates a
CNOT-equivalent operation without additional gate operations,
i.e., for a suitable time $t_C$, $U(t)=e^{-i t_C H_{s}/\hbar}$ becomes
$$
 C_{e} =  CR_z^1(\phi),
 $$
where $R_z^1(\phi)=e^{i\phi\sigma_z^1/2}$ and
\begin{equation}\label{vc}
C = \left ( \begin{array}{cccc}
  1 & 0 & 0 & 0 \\
  0 & 1 & 0 & 0 \\
  0 & 0 & 0 & 1 \\
  0 & 0 & 1 & 0 \\
\end{array}\right ).
\end{equation}

We diagonalise the Hamiltonian $H_{s}=VDV^{\dag}$. Here, $V$
represents the eigenvector matrix
\begin{equation}\label{vr}
V=\left ( \begin{array}{cccc}
  \alpha_{1} & \alpha_2 & 0 & 0 \\
  \beta_1 & \beta_2 & 0 & 0 \\
  0 & 0 & \gamma_1 & \gamma_2 \\
  0 & 0 & \delta_1 & \delta_2 \\
\end{array}\right ),
\end{equation}
and $D$ is the diagonal form of the Hamiltonian, with eigenvalues
$$
\lambda_{1,2}=\hbar\pi[\nu_z^1\pm\sqrt{(\nu_x^2)^2
+(\nu_z^2+\frac{J}{2})^2}]
$$
$$
\lambda_{3,4}=\hbar\pi[-\nu_z^1\pm\sqrt{(\nu_x^2)^2
 +(\nu_z^2-\frac{J}{2})^2}] .
$$
To determine the required evolution time $t_C$, we choose the
target operator as $C_{e}$ and calculate the fidelity
\cite{Jones03},
$$
F(\phi,t) = |Tr[U(t)C_{e}^{\dag}]|/4 .
$$
We find
\begin{equation}\label{fid}
    F(\phi,t)=|e^{-it\lambda_{1}/\hbar}+e^{-it\lambda_{2}/\hbar}-
    \frac{\nu_{x}^{2}e^{i\phi}}{\sqrt{(\nu_{x}^{2})^2+(\nu_{z}^2-\frac{J}{2})^2}}(e^{-it\lambda_3/\hbar}
       -e^{-it\lambda_4/\hbar})|/4\,.
\end{equation}                       
Numerical solutions for $J=215$ s$^{-1}$ are for system {\rm I}
$\phi_C^{\rm I} = 0$, $t_C^{\rm I}=6.15$ ms, resulting in
$F>0.9999$, and for system {\rm II} $ \phi_C^{\rm II} = 0.18\pi$,
$t_C^{\rm II}=4.05$ ms, or $\phi_C^{\rm II}=0.52\pi$, $t_C^{\rm
II}=12.18$ ms, $F>0.999$. Figure \ref{figfid} shows the time
dependence of the fidelity for the three cases.


\subsection{Grover search}

Using CNOT gates, one obtains the controlled phase reversal
$I_{|11\rangle}=W^{2} C W^{2}$ and
$I_{|00\rangle}=e^{i(\pi/2)\sigma_{z}^{1,2}}I_{|11\rangle}$ where
$W^{2}$ denotes the Walsh-Hadamard transform for qubit 2. The
single-qubit gate operations required for the Grover algorithm are
implemented by short radio-frequency pulses, whose duration is
negligible compared to the two-qubit gate. It is therefore
sufficient to consider the coupling to the reservoir during the
evolution under $H_s^{\rm I}$ and $H_s^{\rm II}$.

Including the perturabations, the total Hamiltonians are $H^{\rm
I}_{k}(t)=H^{\rm I}_{s}+\hbar\pi\alpha s_{k}(t)A$ and
    $H^{\rm II}_{k}(t)=H^{\rm II}_{s}+\hbar\pi\alpha s_{k}(t)A$,
 respectively, where $s_{k}(t)$ has been normalized and $\alpha$
 represents the strength of the perturbation.
The perturbed Hamiltonian generates a perturbed phase reversal
$\widetilde{I}_{k,|x\rangle}$, which deviates slightly from the ideal operation $I_{|x\rangle}$,
and which differs for each instance of $s_{k}(t)$.

  The initial state for the Grover search is the uniform superposition
$|\Psi_{0}\rangle=(|00\rangle+|01\rangle+|10\rangle+|11\rangle)/2$
obtained by applying $W^{1,2}$ to $|00\rangle$.
We choose the target state as $|11\rangle$.
Figure  \ref{figgro} shows the sequence of gate operations
for the full Grover search algorithm $G=W^{1,2} I_{|00\rangle}W^{1,2} I_{|11\rangle}$
where we have used $W=W^{-1}$.

In system {\rm I} the a single CNOT gate takes $6.09$ ms, in the
second system either 4.05 or 12.18 ms. To make the duration of the
algorithm in both systems comparable, we replaced the second CNOT
operation in system I by CNOT$^3$, which takes $18.26$ ms.
Since the algorithm includes 2 CNOT gates, the total duration is close to
24 ms in both systems.

In each experiment, we start from the pseudo-pure state
$|\Psi_{0}\rangle$. The perturbed Grover search $G_k=W^{1,2}
\widetilde{I}_{|00\rangle}W^{1,2} \widetilde{I}_{|11\rangle}$
transforms it into $|\Psi_{k}\rangle=G_{k}|\Psi_{0}\rangle$ and the
corresponding density matrix into
$\rho_{k}=|\Psi_{k}\rangle\langle\Psi_{k}|$. Averaging over the
individual signals gives the average (mixed) density matrix
\cite{Ekert96,equivalent}
\begin{equation}\label{reduced}
    \rho=\frac{1}{M}\sum_{k=1}^{M} \rho_{k}.
\end{equation}

\subsection{Decoherence during the search process}\label{purityG}

  To quantify the loss of coherence by the environmental perturbation,
we measured the purity of the Grover search process \cite{purity} by
averaging the purity of the final states for proper input states.
For this purpose, we chose a set of states that is uniformly distributed
over the Bloch sphere.
The uniformly distributed set of input states consists of the 36 states
$|\Psi^{(n)}_{in}\rangle=|\psi_{a}\rangle|\psi_{b}\rangle$, ($a$,
$b=1$, $2$, $\ldots$, $6$) where $|\psi_{a,b}\rangle$ $\in
\{|0\rangle$, $|1\rangle$, $(|0\rangle+|1\rangle)/\sqrt{2}$,
$(|0\rangle-|1\rangle)/\sqrt{2}$, $(|0\rangle+i|1\rangle)/\sqrt{2}$,
$(|0\rangle-i|1\rangle)/\sqrt{2}$ $\}$.

The average purity for the algorithm is then
$P=\frac{1}{36}\sum_{n=1}^{36}Tr\{[\rho^{(n)}]^2\}$ where
$\rho^{(n)}$ denotes the output density matrix after completion of
the quantum search for the input states
$\rho^{(n)}_{in}=|\Psi^{(n)}_{in}\rangle\langle\Psi^{(n)}_{in}|$.
Note that $|\Psi_{0}\rangle$ is one of the 36 input states.



When the systems are embedded in the reservoirs shown in Figure
\ref{fig1}, the final states $\rho^{(n)}$ can be calculated by
solving the Bloch- Redfield equations.
In the eigenbase of the
system Hamiltonian $H_s$ in Eq. (\ref{Hrs}), the Bloch- Redfield
equations are
\begin{equation}\label{Redf}
    \dot{\rho}_{nm}=-i\omega_{nm}\rho_{nm}-\sum_{k,l}R_{nmkl}\rho_{kl}.
\end{equation}
where $R_{nmkl}$ denotes the partial decoherence rates
\begin{equation}\label{Relax}
    R_{nmkl}=\delta_{lm}\sum_{r}\Lambda_{nrrk}+\delta_{nk}\sum_{r}\Lambda^{*}_{lrrm}
    -\Lambda_{lmnk}-\Lambda^{*}_{knml}.
\end{equation}
$\Lambda_{lmnk}$ denotes the element of the relaxation tensor. Eq.
(\ref{Relax}) shows that one can obtain $R_{nmkl}$ through the
real parts of the relaxation tensor represented as
\begin{equation}\label{tensor}
    Re\{\Lambda_{lmnk}\}=\frac{1}{4\pi} S(\omega_{nk})
    (A_{lm}A_{nk}) .
\end{equation}
When $A=\sigma^{2}_{z}$, it takes the form 
\begin{equation}\label{a2}
    A^{I}=\left (\begin{array}{cccc}
      0 &   0 & -1 & 0 \\
      0 & -0.6606 & 0 & 0.7507 \\
      -1 & 0 & 0 & 0 \\
      0 & 0.7507 & 0 & 0.6606 \\
    \end{array}
    \right ),
\hspace{1.0 cm}
    A^{II}=\left (\begin{array}{cccc}
      -0.8695 &   0 & 0 & 0.4940 \\
      0 & 0 & -1 & 0\\
      0 & -1 & 0 & 0 \\
      0.4940 & 0 & 0 & 0.8695 \\
    \end{array}
    \right )
\end{equation}
in the energy representation of the systems {\rm I} and {\rm II}, respectively.
When $A=\sigma^{1}_{z}$, it is diagonal because $[\sigma^{1}_{z},H_{s}]=0$.
The matrices in the two systems are then
\begin{equation}\label{a1}
    A^{I}=\left (\begin{array}{cccc}
      -1 &   0 & 0 & 0 \\
      0 & 1 & 0 & 0 \\
      0 & 0 & -1 & 0 \\
      0 & 0 & 0 & 1 \\
    \end{array}
    \right ),
\hspace{1.0 cm}
    A^{II}=\left (\begin{array}{cccc}
      1 &   0 & 0 & 0 \\
      0 & -1 & 0 & 0\\
      0 & 0 & -1 & 0 \\
      0 & 0 & 0 & 1 \\
    \end{array}
    \right ).
\end{equation}

Eqs. (\ref{Redf}-\ref{tensor}) show that decoherence can be suppressed by choosing the
parameters of $H_s$ such that the elements of the tensor ${\bf R}$ get small
\cite{Grigorenko}.
Qualitatively, the influence of the environment
depends on the size of $|S(\omega_{nk})A_{nk}|$.

\subsection {Numerical evaluation}

We first present a numerical evaluation of Eq. (\ref{Redf})
to calculate the purity $P_{1,2}$ for the two systems.
Figures \ref{figpurityH2} and \ref{figpurityC1} summarize the
result.
Figures \ref{figpurityH2} (a-c) show the purity for the
perturbation operator $A=\sigma^{2}_{z}$ and the reservoirs R1-R3.
In Figure (a), the increase of $\alpha$, has very little effect on
$P_2$ because $S(\omega^{\rm II}_{nk})\approx0$. However, $P_1$
decreases significantly, because $S(\omega^{\rm I}_{42})A^{\rm
I}_{42}=0.7507$. Comparing the two systems in R1, one finds that
system {\rm II} is more robust.
In Figure (\ref{figpurityH2} b) the situation is reversed and system {\rm I}
is more robust.
In Figure (\ref{figpurityH2} c), $P_1$ decreases faster than $P_2$ because
$|S(\omega^{\rm I}_{31})A^{\rm I}_{31}|>|S(\omega^{\rm
II}_{41})A^{\rm II}_{41}|$.
These results illustrate the possibility to suppress decoherence
by choosing an appropriate energy level structure.

When $A=\sigma^{1}_{z}$, it is diagonal in the eigenbase of the Hamiltonian.
In this case, the reservoir does not induce transitions, but only causes dephasing,
according to Eq. (\ref{a1}).
The energy level structure has then only a small effect on the decoherence rates.
>From Eq. (\ref{tensor}), one finds that the
reservoir affects the quantum system only through its static part $S(0)$.
Figure \ref{figpurityC1} shows the resulting purities $P_1$ and $P_2$
for the case that qubit 1 is coupled to the reservoirs R3-R4.
In Figure (a) both $P_1$ and $P_2$ remain close to
$1$ because $S(0)\approx 0$.
In Figure (b), however, $S(0) = 1$, and both systems are affected
in a similar way.

\section{Experimental procedure and results} 

\subsection{Implementation of Hamiltonians}

For the experimental implementation, we chose Carbon-13 labelled chloroform
(CHCl$_3$) dissolved in d6-acetone as the quantum register.
We chose the carbon as qubit 1 and the proton as qubit 2.
The noise term in
Eq. (\ref{Hen}) is introduced by an offset variation of the
transmitter on channel 2.
The Hamiltonian of the two qubit NMR system is thus
\begin{equation}\label{NMR}
    H_{NMR,k}(t)=\frac{1}{2}\hbar\omega^{1}_{z}\sigma_{z}^{1}
   +\frac{1}{2}\hbar\omega^{2}_{z}\sigma_{z}^{2}+
   \frac{1}{2}\hbar\pi J\sigma_{z}^{1}\sigma_{z}^{2}+
   \hbar\pi s_{k}(t)A,
\end{equation}
where $A=\sigma_{z}^{2}$ or $\sigma_{z}^{1}$ and the coupling is
$J=215$ Hz.

The transverse field
in Eq. (\ref{Hrs}) is applied as a radio frequency (rf) field, which can be written as
\begin{equation}\label{rf}
    H_{rf}=-\frac{1}{2}\hbar\omega^{2}_{x}\sigma_{x}^{2}
\end{equation}
in the rotating reference frame.

In the experimental implementation, the perturbation and thus the
total Hamiltonian are piecewise constant for short periods $\tau =
304.38$ $\mu$s. For each of these periods, we realized the total
evolution $e^{-i\tau H_{k}(t)/\hbar}$ as $e^{-i\tau H_{NMR,k}/\hbar}
e^{-i\tau H_{rf}/\hbar}$, i.e. by a short free precession period
followed by a small flip-angle pulse \cite{Vandersypen}. This is a
good approximation
when $\tau\ll 2\pi/\omega_{nm}$. 

\begin{figure}
\includegraphics[width=6in]{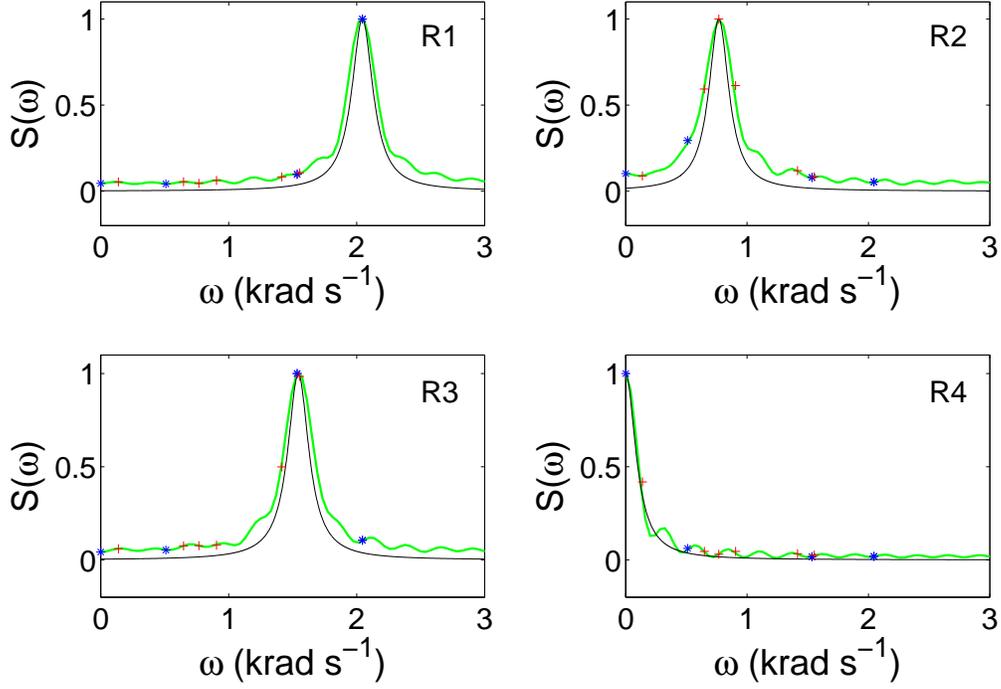}
\caption{(Color online) Normalized spectral functions for the four
reservoirs (R1-R4) with $\Gamma=100$ s$^{-1}$  and
$\omega_{0}=\omega^{I}_{42}$, $\omega^{II}_{32}$,
$(\omega^{I}_{31}+\omega^{II}_{41})/2
\approx\omega^{I}_{31}\approx\omega^{II}_{41}$ and $0$,
respectively. The thick curves represent the generated spectral
functions for each reservoir, which were generated by averaging over
$2500$ noise signals. The thin curves represent the corresponding
theoretical spectral functions. The transition angular frequencies
of the systems \rm {I} and \rm {II} are marked by "*" and "+". }
\label{fig1}
\end{figure}

\begin{figure}
\includegraphics[width=5in]{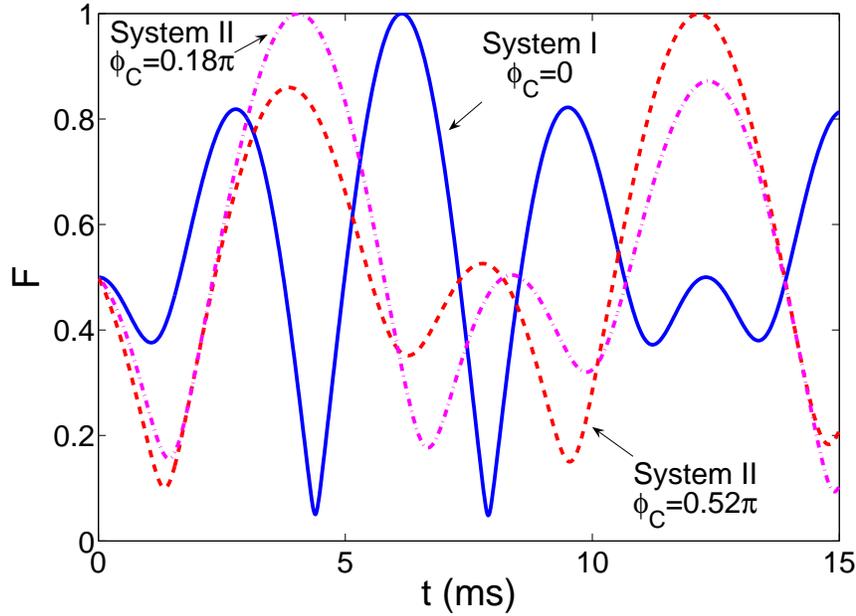}
\caption{(Color online) Time dependence of the fidelity for the
CNOT-equivalent operation. In system {\rm I}, shown as the solid
curve, the proper evolution time is $6.15$ ms, and $\phi_C=0$. In
system {\rm II}, shown as the dash-dotted and dashed curves, the
proper evolution times are $4.05$ ms and $12.18$ ms for
$\phi_C=0.18\pi$ and $0.52\pi$, respectively.} \label{figfid}
\end{figure}

\begin{figure}
\includegraphics[width=5in]{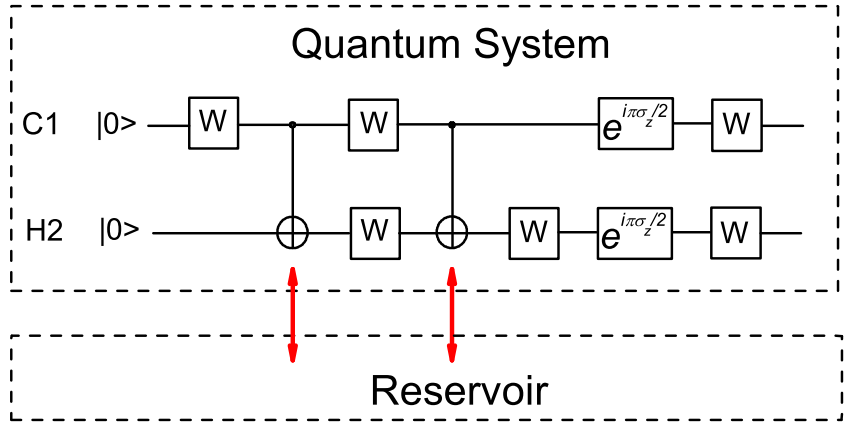}
\caption{(Color online) Gate sequence for the Grover search in an
open quantum system. The time order is from left to right. $W$
denotes the Walsh-Hadamard transform, and the red arrows denote the
interaction between the system and reservoir. The duration of the
gates $W$ and $e^{i\pi\sigma_{z}/2}$ is so short that it can be
ignored. } \label{figgro}
\end{figure}

\begin{figure}
\includegraphics[width=7in]{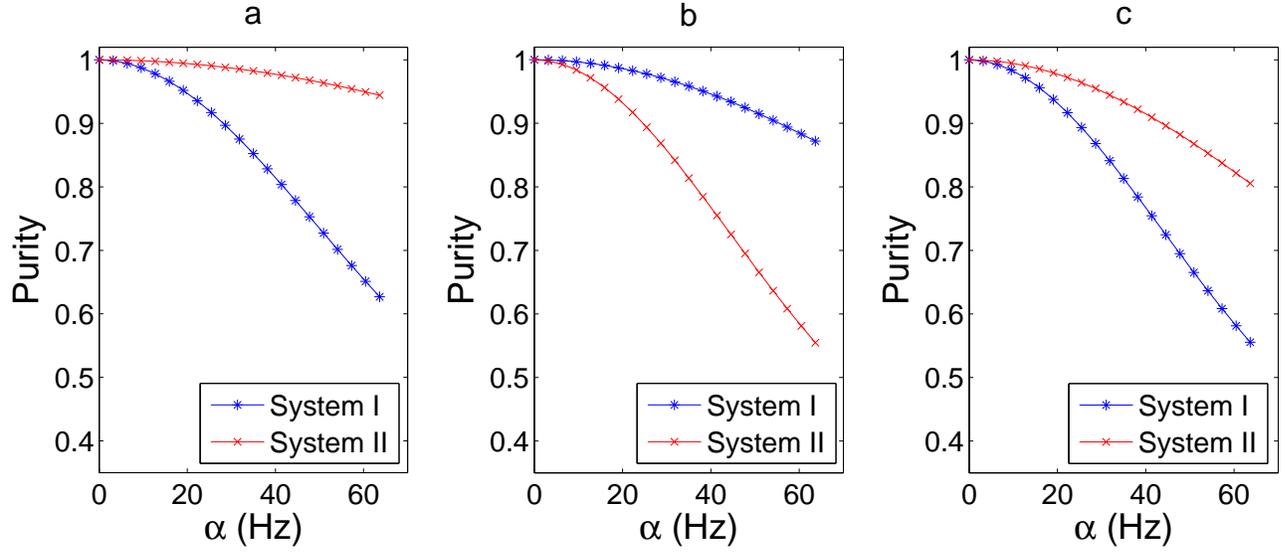}
\caption{(Color online) Purity of Grover search process as a
function of the coupling strength to the environment when qubit 2 is
coupled to the reservoirs in R1-R3, respectively. The data points
obtained in systems {\rm I} and {\rm II} are marked by "*" and
"$\times$", respectively. }\label{figpurityH2}
\end{figure}

\begin{figure}
\includegraphics[width=5in]{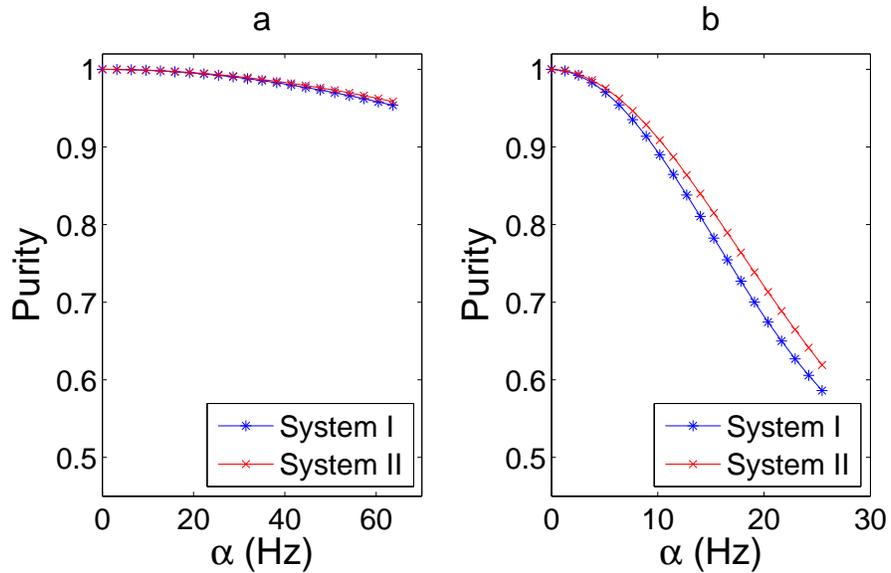}
\caption{(Color online) Same as Fig. \ref{figpurityH2}, but for
qubit 1 coupled to reservoirs R3-R4.}\label{figpurityC1}
\end{figure}

For the experimental implementation of the reservoirs R1-R4 shown in
Figure \ref{fig1}, we used $M=12$ noise traces. The corresponding
spectral functions are similar to those represented in Figures
\ref{fig1} (a-d).

\begin{figure}
\includegraphics[width=6.5in]{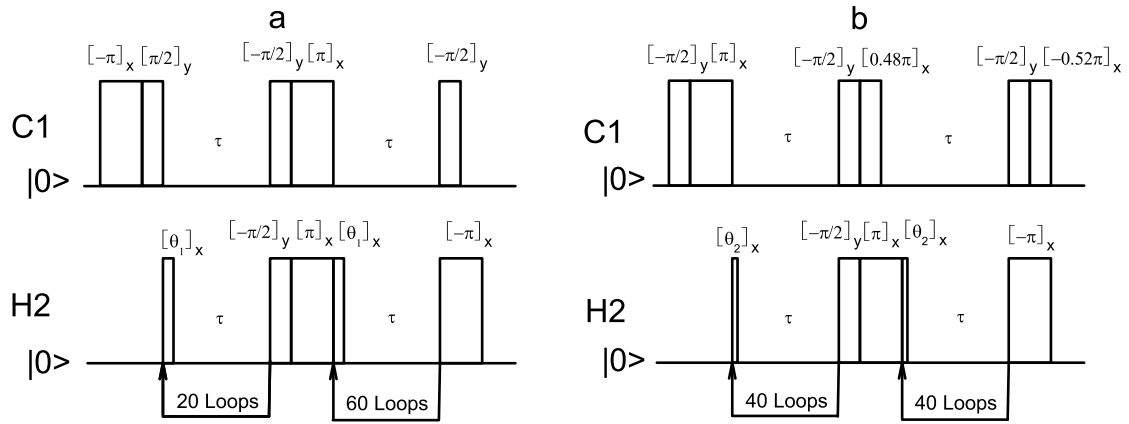}
\caption{Pulse sequences to implement the Grover search in systems
{\rm I} shown in (a) and {\rm II} shown in (b). The flip angles
$\theta_{1}$ and $\theta_{2}$ for the two systems differ by a factor
of 2. During the delay denoted by $\tau$, the systems evolute under
the natural Hamiltonian and the noise signal of Eq. (\ref{NMR}). }
\label{figpul}
\end{figure}

  The experiments start with the effective pure state $|00\rangle\langle00|$
prepared by spatial averaging \cite{s11,dujf}.
The pulse sequence
$[\alpha]_{x}^{2}-[grad]_{z}-[\pi/4]_{x}^{2}-\frac{1}{4J}-[\pi]_{x}^{1,2}-\frac{1}{4J}-[-\pi]_{x}^{1,2}
-[-\pi/4]_{y}^{2}-[grad]_{z}$ transforms the system from the
equilibrium state to the effective pure state
$|00\rangle\langle00|$.
Here
$\alpha=\arccos(2\gamma_{1}/\gamma_{2})\approx \pi/3$, where
$\gamma_{1}$ and $\gamma_{2}$ denote the gyromagnetic ratios of
$^{13}$C and $^1$H, respectively, and
 $[grad]_{z}$ denotes a gradient pulse along the $z$-axis.
$\frac{1}{4J}$ denotes the evolution caused by $H_{NMR}$ for a time
$\frac{1}{4J}$. The pulses are applied from left to right. The
complete pulse sequences for the implementation of the Grover search
in the two systems are shown as Figures \ref{figpul} (a) and (b),
respectively.
\subsection{Grover Search in systems not coupled to reservoirs}

\begin{figure}
\includegraphics[width=4in]{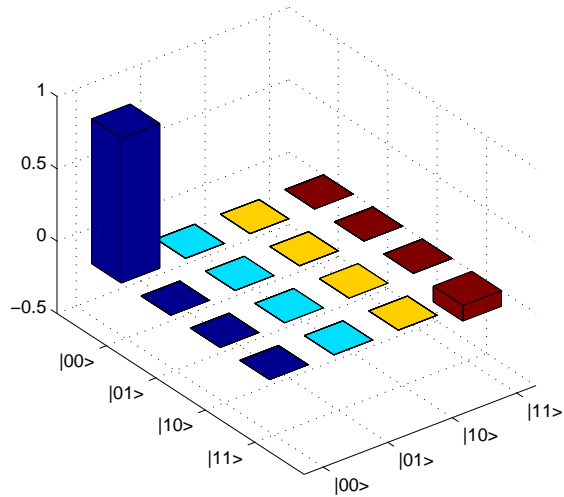}
\caption{(Color online) Experimentally measured density matrix when
the system lies in the initial pseudo-pure state
$|00\rangle\langle00|$.} \label{figini}
\end{figure}

When the system is in $|00\rangle\langle00|$, we experimentally
measured the density matrix shown as Figure \ref{figini} through
state tomography \cite{ChuangPRSL}. In either system the Grover
algorithm is repeated $12$ times, using different noise traces.
The target state is $|11\rangle$.
The final NMR signals are obtained by summing the
$12$ signals acquired via the readout pulse.

\begin{figure}
\includegraphics[width=7in]{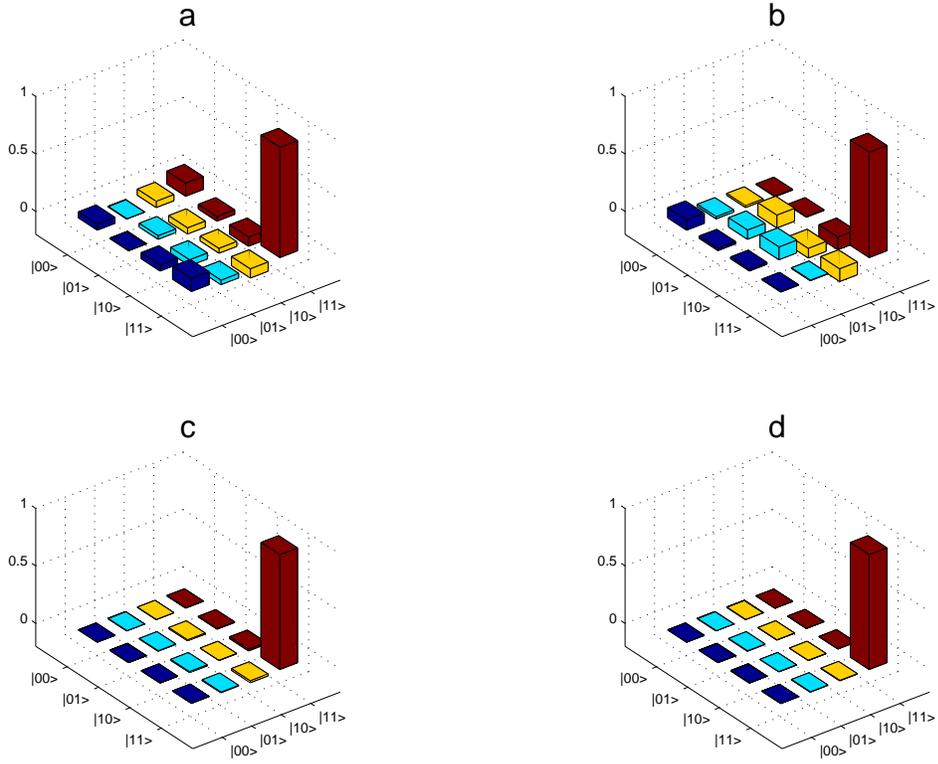}
\caption{(Color online) Experimentally measured density matrices
after the completion of the Grover search in the systems {\rm I}
[shown as (a)], and {\rm II} [shown as (b)] when the reservoir is
not applied. The target state is chosen as $|11\rangle$. The
matrices have been normalized. Only the real parts of the elements
are plotted. The imaginary parts are less than $18\%$. In order to
estimate the errors caused by the imperfections of rf pulses and
natural decoherence, Figures (c-d) show the simulated results
obtained by NMR simulator where the rf pulses are perfect and no
natural decoherence exists, corresponding to Figures (a-b),
respectively. } \label{figno}
\end{figure}

We first implement the Grover search in the systems without
engineered noise, i.e., the noise signal $\alpha s_k(t)$ ($k=1$,
$2$, $\cdots$, $12$) is not applied to the quantum systems. After
the completion of the search algorithm, the density matrices of the
two systems are shown as Figures \ref{figno} (a-b), respectively.

In the experiments, the imperfections of the rf pulses and natural
decoherence cause errors in the search results. In order to
distinguish these errors from those that are due to the engineered
"noise" that we investigate here, we use the results of in Figures
\ref{figno} (a-b) as the references for subsequent experiments in
the systems coupled to the reservoirs. We denote these reference
states as $\rho^{\rm I}_{0}$ and $\rho^{\rm II}_{0}$.

To estimate the effects of the errors caused by imperfections of the
rf pulses and natural decoherence, we compare the experimental
results to simulation data, where the rf pulses are perfect and no
natural decoherence exists. These results are shown as Figures
\ref{figno} (c-d) corresponding to (a-b), respectively. The overlap
between $\rho_{0}$ and its corresponding simulated result is 0.97
for system {\rm I} and 0.91 for system {\rm II}.

\subsection{Search results in systems coupled to reservoirs}\label{Reserv2}

To simulate the noisy reservoir and observe the decoherence effect,
we apply $M=12$ different noise traces $\alpha s_k(t)$ to the system
during the implementation of the Grover search and add the resulting signals.
We quantify the resulting decoherence by the fidelity
$F_{\rho}=Tr(\rho_{0} \rho)$.

\begin{figure}
\includegraphics[width=7in]{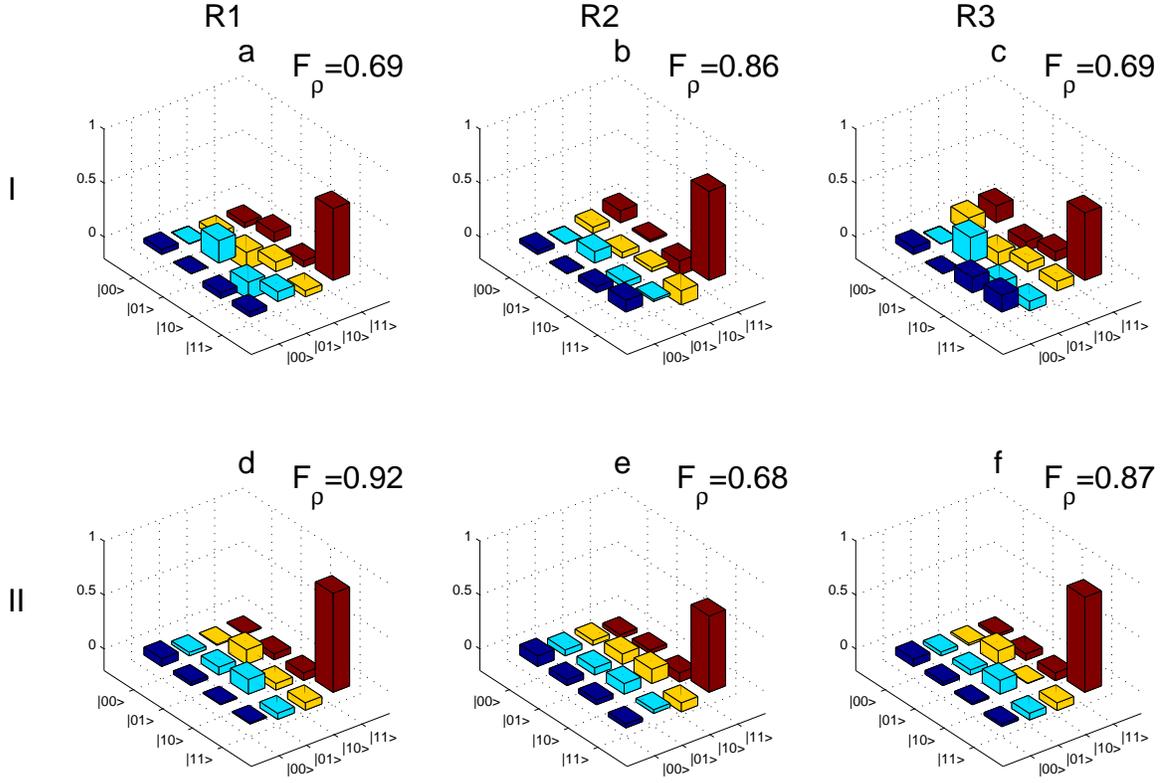} 
\caption{(Color online) Experimentally measured density matrices
after the completion of the Grover search when qubit $2$ is coupled
to reservoirs R1-R3, shown as the three columns from left to right.
The two rows of figures show the results obtained in systems {\rm I}
and {\rm II}, respectively. $F_{\rho}$ denotes the fidelity of the
search result with respect to the corresponding noiseless result.}
\label{figph}
\end{figure}

Figure \ref{figph} shows the resulting density operators for the
case where the coupling operator is $A=\sigma^{2}_{z}$ and the
reservoir is R1-R3. The upper row corresponds to Fig. I, the lower
row to Fig. II. In all cases, the coupling strength was set to
$\alpha=63.66$ Hz. The fidelity $F_{\rho}$ shown in the figures was
calculated as the overlap between the states resulting from the
noisy experiment and those from the experiment without the
reservoir.

The observed results are in good agreement with the predictions from
the numerical simulations shown in Figure \ref{figpurityH2}. For
example, the reduction of $P_1$ in Figure \ref{figpurityH2} (a)
leads to the low fidelity $F_{\rho}=0.69$ in Figure \ref{figph} (a),
while $F_{\rho}$ in Figure \ref{figph} (d) is $0.92$ and $P_2$ in
Figure \ref{figpurityH2} (a) remains close to $1$. In Figures
\ref{figph} (c) and (f), the fidelity in system {\rm II} is larger
than that in system {\rm I}, in good agreement with the result that
$P_{2}>P_{1}$ in Figure \ref{figpurityH2} (c).

\begin{figure}
\includegraphics[width=7in]{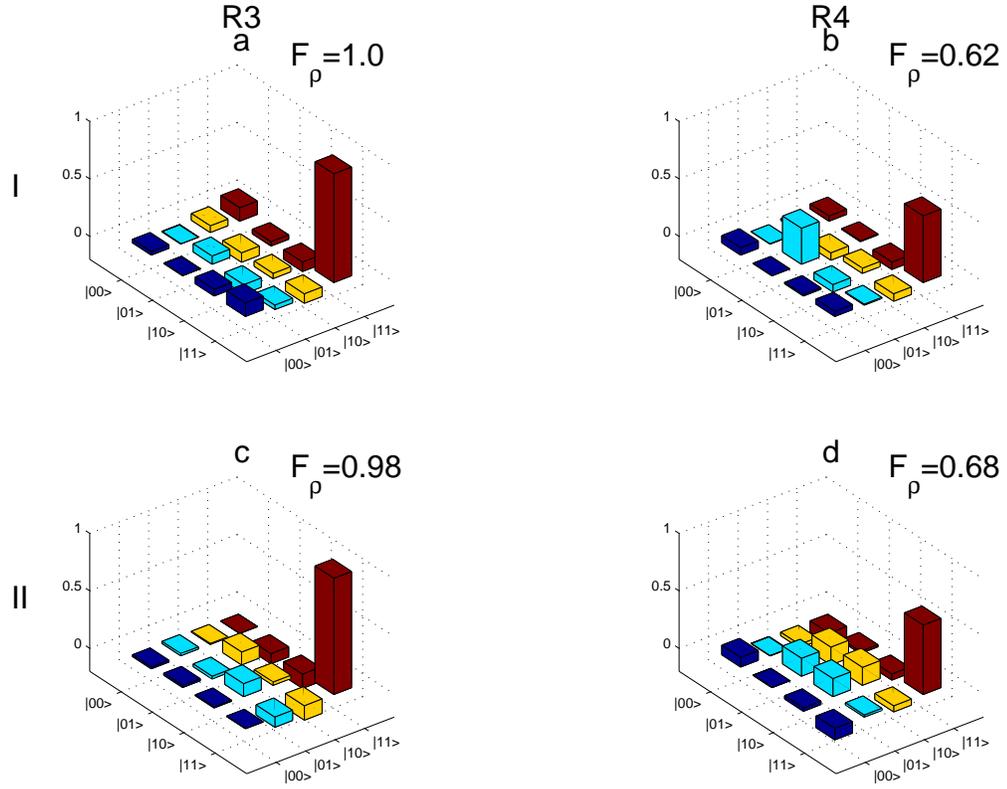}    
\caption{(Color online) Experimentally measured density matrices
after the completion of the Grover search when qubit $1$ is coupled
to reservoirs R3-R4, shown as the two columns from left to right. }
\label{figpc}
\end{figure}

Figure \ref{figpc} shows the search result for the case where the
coupling operator is $A=\sigma^{1}_{z}$ and the systems are coupled
to reservoirs R3-R4. In Figures (a) and (c), $\alpha=63.66$ Hz; in
Figures (b) and (d), $\alpha=25.46$ Hz. The experimental results
agree with the results in Figure \ref{figpurityC1}. The much higher
fidelity of the first column show clearly that the reservoir affects
the system only through $S(0)$.




\section{Generalizations}

The above description of the decoherence process uses the semiclassical approximation,
where the environment interacts with the system through classical fields.
The results are easily generalized to the case of a quantum mechanical environment.
For this purpose, we describe the total system (quantum register plus bath)
by the Hamiltonian
\begin{equation}\label{quan_env}
   H_{tot}= H_{s}+H_{B}+H_{I}
\end{equation}
where $H_{B}$ denotes the Hamiltonian of the bath, and $H_{I}$
denotes the coupling between the system and the bath. For the
purpose of comparison we choose $H_{I}=AX$ where $X$ denotes an
operator of the bath.

In the quantum mechanical description, the dynamics of the quantum
register are obtained by tracing over the degrees of freedom of the
environment. It is thus possible to recover the Bloch- Redfield
equations. In the eigenbase of $H_{s}$, the rates $\Lambda_{lmnk}$
(\ref{Redf}-\ref{tensor}) become \cite{Srorcz,Srorcz2}
\begin{equation} \label{ten_quant}
\Lambda_{lmnk} = A_{lm} A_{nk} \int_{0}^{\infty} \frac{1}{\hbar^{2}}
e^{-i\omega_{nk}t}\langle X(t)X(0)\rangle dt
\end{equation}
where $X(t)=e^{iH_{B}t/\hbar}Xe^{-iH_{B}t/\hbar}$. The brackets
$\langle \ldots \rangle$ denote the thermal average over the bath
degrees of freedom. The Fourier transform of this correlation
function corresponds to the spectral function $S(\omega_{nk})$ in
Eq. (\ref{tensor}). This means that the fully quantum model
described by Eq. (\ref{quan_env}) can be mapped to the quantum
system under classical noise described by Eq. (\ref{Hen})
\cite{equivalent}. Consequently our results are equally applicable
to the fully quantum-mechanical case.


 Besides the coupling between the quantum system and its
environment, pulse imperfections (i.e. nonideal gate operations)
also induce decoherence.
Our results can be
easily generalized to investigate the effect of the pulse
imperfections on the decoherence rate. The generalization is
illustrated by rewriting Eq. (\ref{Hen}) as
\begin{equation}\label{imperfect}
   H(t)=\frac{1}{2}\hbar \left[ \Omega_{z}^{1}(t) \sigma_{z}^{1}
   + \omega^{2}_{z}\sigma_{z}^{2}
   - \omega^{2}_{x}\sigma_{x}^{2}+
  \pi J\sigma_{z}^{1}\sigma_{z}^{2} \right]
\end{equation}
where $\Omega_{z}^{1}(t)=\omega^{1}_{z}+2\pi s(t)$, when
$A=\sigma_{z}^{1}$. $\Omega_{z}^{1}(t)$ denotes the strength of the
pulse that randomly fluctuates about $\omega^{1}_{z}$, and $2\pi
s(t)$ describes the fluctuation. Using our methods, one can discuss
the effects of different fluctuations and search for experimental conditions
that minimize the effect of pulse imperfections.

\section{conclusion}

We have investigated, experimentally and theoretically, the effect of
different reservoirs on the decoherence of quantum registers
during the execution of a quantum algorithm.
While we have used a semiclassical system for these investigations,
the results are easily adapted to a quantum mechanical environment,
such as a spin-boson model.

The system Hamiltonian as well as the coupling operator determine
whether the environment causes pure dephasing or also induces
transitions. The situation that is probably most relevant for
quantum information processing is the case where the coupling
operator is diagonal in the eigenbase of the system Hamiltonian. In
this case, the environment causes pure dephasing and only the static
part of the perturbation, $\propto |S(0)|$ causes decoherence.

While we have chosen a 2-qubit system for this investigation, the
results are completely general and can be applied directly to
multi-qubit systems.
It is possible to use this method of simulating dissipative quantum systems
for related phenomena, such as dissipative quantum phase transitions
\cite{Capriotti}.
In the field of quantum information processing, our results indicate possible
ways for suppressing decoherence.

\section{Acknowledgment}

We thank Prof. Jiangfeng Du and Dr. Bo Chong for helpful
discussions. The experiments were performed at the Interdisciplinary
Center for Magnetic Resonance. This work is supported by the
Alexander von Humboldt Foundation, the National Natural Science
Foundation of China under grant No. 10374010, the DFG through Su
192/19-1, and the Graduiertenkolleg No. 726.



\end{document}